\documentclass[reprint,superscriptaddress,aps,prl]{revtex4-2}
\usepackage[T1]{fontenc}

\begin{document}
	\title{Nuclear electric resonance}
	
	\author{Jian Leng}
	\affiliation{ State Key Laboratory of Low Dimensional Quantum Physics, Department of Physics, \\ Tsinghua University, Beijing 100084, China}
	\author{Fan Yang}
	\affiliation{ State Key Laboratory of Low Dimensional Quantum Physics, Department of Physics, \\ Tsinghua University, Beijing 100084, China}
	\author{Xiang-Bin Wang}
	\email{ xbwang@mail.tsinghua.edu.cn}
	\affiliation{ State Key Laboratory of Low Dimensional Quantum Physics, Department of Physics, \\ Tsinghua University, Beijing 100084, China}
	\affiliation{ Jinan Institute of Quantum technology, SAICT, Jinan 250101, China}
	\affiliation{ Shanghai Branch, CAS Center for Excellence and Synergetic Innovation Center in Quantum Information and Quantum Physics, University of Science and Technology of China, Shanghai 201315, China}
	\affiliation{ Shenzhen Institute for Quantum Science and Engineering, and Physics Department, Southern University of Science and Technology, Shenzhen 518055, China}
	\affiliation{ Frontier Science Center for Quantum Information, Beijing, China}
	
	\begin{abstract}
		Nuclear-spin qubits have long coherence time and are desirably applied into quantum information processing. However, the existing methods either fail to address single nucleus (such as nuclear magnetic resonance), or severely affect nuclear coherence time (such as electrical nuclear manipulation based on hyperfine stark effect, ENMHSE). Here we propose an electrical nuclear manipulation called nuclear electric resonance which can on the one hand address the single nuclear qubit, and on the other keep the long coherence time. Applying this, we construct universal quantum gates with external electric field. These universal gates are practicable for arbitrary $S\ge 1$ spin nuclei. Given the much longer coherence time of nuclear electric resonance, we improve the number of single-qubit operations by three orders of magnitude compared with that of ENMHSE.
	\end{abstract}
	
	
	\maketitle

	\section{I. Introduction}
	
	Quantum computing~\cite{divincenzo1995quantum,ekert1996,preskill2018,arute2019quantum,wu2021strong,zhong2021phase} can solve some important complex problems which go beyond the power of classical information processing. Decoherence is the biggest barrier to realize a quantum computer in the real world. Nuclear spin is an important candidate for quantum computing because of its long coherence time~\cite{Nielsen2002,park2017second,sharma2019enhancement,serafin2021nuclear}. Naturally, we may try to manipulate nuclear spins by magnetic field. However, this method such as nuclear magnetic resonance is technically difficult to modulate the magnetic field exactly in micro areas and therefore it only manipulates ensemble states rather than a single quantum state, though it can have a long coherence time. This makes it difficult to design a scalable quantum computer. Electric field (EF) can be tuned exactly in micro areas by existing matured technology, and some electrical nuclear manipulation proposals~\cite{thiele2014electrically,laucht2015electrically,sigillito2017all} based on hyperfine stark effect (ENMHSE) are presented. But hyperfine stark effect affects coherence time since it is the mutual coupling between nuclear spin and electron angular momentum. Remarkably, it has been experimentally demonstrated~\cite{asaad2020coherent} that EF can couple with single nucleus based on electric quadrupole effect (EQE)~\cite{Bloembergen1961,Slichter2013}. EQE holds a long coherence time since nuclear spin is unilaterally driven by electric field gradient (EFG) produced by electrons. This EFG is an `external' field for nucleus. However, the existing study~\cite{asaad2020coherent,Slichter2013} of EQE is essentially based on the model of static electric field, while there is no study on manipulating nuclear states. Here we present a general theoretical model for nuclear EQE with time-dependent electromagnetic field (EMF). With the model we systematically shows how to realize nuclear electric resonance (NER) which can precisely manipulate a single nucleus by oscillating EF. Universal quantum gates are achieved by NER with nuclear interaction control. Our NER quantum computing proposal has huge advantages in application: it can be applied to arbitrary $S\ge 1$ spin nuclei and has a remarkably excellent performance on single-qubit operation since its coherence time is much longer than ENMHSE's.
	
	\section{II. Nuclear Electric Quadrupole Effect in electromagnetic Field}
	
	Consider a nucleus at the origin forced by vacuum EMF. We take scalar potential $\phi({\bf r},t)=0$~\cite{Scully1997}, and vector potential $A_\alpha({\bf r},t)(\alpha=x,~y,~z)$ to first order
	\begin{equation}
		A_\alpha({\bf r},t)=A_\alpha({\bf 0},t)+\partial_\beta A_\alpha({\bf 0},t)r_\beta,
	\end{equation}
	with the Einstein summation convention, i.e., when an index variable appears twice in a single term it implies summation over $\{x,~y,~z\}$. Consider a scalar function
	\begin{equation}
		G({\bf r},t)\equiv A_\alpha({\bf 0},t)r_\alpha+\frac{1}{2}\partial_\alpha A_\beta({\bf 0},t)r_\alpha r_\beta.
	\end{equation}
	Taking the gauge transformation
	\begin{eqnarray}
		A^\prime_\alpha({\bf r},t)=&&A_\alpha({\bf r},t)-\partial_\alpha G({\bf r},t),\nonumber\\
		\phi^\prime({\bf r},t)=&&\phi({\bf r},t)+\partial_tG({\bf r},t),
	\end{eqnarray}
	and meanwhile noticing that
	\begin{eqnarray}
		\partial_\alpha A_\beta({\bf 0},t)-\partial_\beta A_\alpha({\bf 0},t)=&&\varepsilon_{\alpha\beta\gamma}B_\gamma({\bf 0},t),\nonumber\\
		-\partial_tA_\alpha({\bf 0},t)=-\partial_tA_\alpha({\bf 0},t)-&&\partial_\alpha\phi({\bf 0},t)=E_\alpha({\bf 0},t),
	\end{eqnarray}
	where $\varepsilon_{\alpha ij}$ is the Levi-Civita symbol, we obtain
	\begin{eqnarray}
		A^\prime_\alpha({\bf r},t)=&&-\frac{1}{2}\varepsilon_{\alpha\beta\gamma}r_\beta B_\gamma({\bf 0},t),\nonumber\\
		\phi^\prime({\bf r},t)=&&-E_\alpha({\bf 0},t)r_\alpha-\frac{1}{2}\partial_\alpha E_\beta({\bf 0},t)r_\alpha r_\beta.
		\label{potential}
	\end{eqnarray}
	With this we can write down the Hamiltonian for our system, a nucleus under the time-dependent EMF
	\begin{eqnarray}
		&&~~\mathcal{H}=\sum_z^Z\Big[\frac{1}{2m^{(z)}}(p_\alpha^{(z)}-eA^\prime_\alpha({\bf r}^{(z)},t))(p_\alpha^{(z)}-eA^\prime_\alpha({\bf r}^{(z)},t))\nonumber\\
		&&+e\phi^\prime({\bf r}^{(z)},t)+V({\bf r}^{(z)})+\gamma^{(z)}s_\alpha^{(z)}B_\alpha({\bf 0},t)\Big]\nonumber\\
		&&+\sum_n^N\left[\frac{1}{2m^{(n)}}p_\alpha^{(n)}p_\alpha^{(n)}+V({\bf r}^{(n)})+\gamma^{(n)}s_\alpha^{(n)}B_\alpha({\bf 0},t)\right],\nonumber\\
		\label{H_1}
	\end{eqnarray}
	where $V({\bf r})=-\frac{V_0}{1+\exp(\frac{|{\bf r}|-R_0}{a})}$ is the Woods-Saxon potential that describes the nuclear forces applied on each nucleon, index $(z)$ and $(n)$ represent protons and neutrons respectively, $Z$ and $N$ are the number of protons and neutrons respectively, $m$ is the mass, $e$ is the elementary charge, $\gamma$ is the gyromagnetic ratio, $s$ is the spin, $R_0$ is the nuclear radius, $V_0$ and $a$ are constants. Applying Eq.~(\ref{potential}) into Eq.~(\ref{H_1}) we obtain
	\begin{widetext}
		\begin{eqnarray}
			\mathcal{H}=&&\sum_{z,n}^{Z,N}\left[-\frac{e}{2m^{(z)}}\varepsilon_{\alpha\beta\gamma}r_\beta 	p_\gamma+\gamma^{(z)}s_\alpha^{(z)}+\gamma^{(n)}s_\alpha^{(n)}\right]B_\alpha({\bf 0},t)+\sum_z^Z\frac{e^2}{8m^{(z)}}\Big[r_\alpha^{(z)}r_\alpha^{(z)}B_\beta({\bf 0},t)B_\beta({\bf 0},t)-r_\alpha^{(z)}r_\beta^{(z)}B_\alpha({\bf 0},t)B_\beta({\bf 0},t)\Big]\nonumber\\
			&&+\sum_{z,n}^{Z,N}\left[\frac{1}{2m^{(z)}}p_\alpha^{(z)}p_\alpha^{(z)}+V({\bf 	r}^{(z)})+\frac{1}{2m^{(n)}}p_\alpha^{(n)}p_\alpha^{(n)}+V({\bf r}^{(n)})\right]+\sum_z^Z\left[-eE_\alpha({\bf 0},t)r_\alpha^{(z)}\right]-\sum_z^Z\left[\frac{e}{2}\partial_\alpha E_\beta({\bf 0},t)r_\alpha^{(z)}r_\beta^{(z)}\right].
			\label{H_2}
		\end{eqnarray}
	\end{widetext}
	Consider the terms in right hand side of Eq.~(\ref{H_2}) one by one. The first summation is simply the interaction between nuclear magnetic moment and magnetic field, hence equivalent to $\gamma_n S_\alpha B_\alpha({\bf 0},t)$ where $\gamma_n$ is the nuclear gyromagnetic ratio and $S$ is the nuclear spin. Compared with this, the second summation can be ignored since it is a second order small quantity. By the Wigner-Eckart theorem the third summation is omitted since it is a zero-order spherical harmonic tensor operator and equivalent to a constant. The forth summation is also negligible because nucleus is so stable that we only concern about the protons move around the origin. Applying the Wigner-Eckart theorem and the Gauss's law $\partial_\alpha E_\alpha({\bf 0},t)=0$ we can rewrite the fifth summation into the following equivalent form after some tedious calculations:
	\begin{equation}
		\mathcal{H}_\mathrm{Q}\equiv-\frac{1}{2}\tilde{Q}\partial_\alpha E_\beta({\bf 0},t)(S_\alpha S_\beta+S_\beta S_\alpha),
		\label{HQ_final}
	\end{equation}
	where $\tilde{Q}=\frac{eQ}{2S(2S-1)\hbar^2}$ and $Q\equiv Z\langle SS|3z^2-r^2|SS\rangle$	is the nuclear electric quadrupole momentum. Eq.~(\ref{HQ_final}) is the general form of EQE in EMF. Given the discussion above, Eq.~(\ref{H_2}) can be written more neatly as
	\begin{equation}
		\mathcal{H}=\gamma_n S_\alpha B_\alpha({\bf 0},t)+\mathcal{H}_\mathrm{Q}.
		\label{H_final}
	\end{equation}
	This is our major equation for Hamiltonian. To confirm the correctness of Eq.~(\ref{H_final}), we suppose magnetic field at the origin is independent with time and all components of electric field have the same frequency and phase, it turns to the Hamiltonian in Ref.~\cite{asaad2020coherent}. If we only consider the time-independent electric field, Eq.~(\ref{H_final}) will be the same as EQE in a static electric field presented in Ref.~\cite{Slichter2013}.
	
	\section{III. Nuclear Electric Resonance}
	
	Nuclear electric quadrupole momentum $Q$ is extremely small, about $10^{-26}\sim10^{-28}\mathrm{m}^2$. Straightly using Eq.~(\ref{HQ_final}), we need a huge EFG which is impossible for the existing technology. To solve this problem, we use external oscillating EF to manipulate atomic electron states and they produce a sufficiently strong EFG on the nucleus. We can precisely manipulate nuclear state by controlling oscillating EF, called nuclear electric resonance.
	
	The Hamiltonion for an electron in an atom with static magnetic field $(0,~0,~B_0)$ is
	\begin{equation}
		\mathcal{H}_0=\frac{{\bf p}_e^2}{2m_e}-\frac{kZe^2}{r}+\gamma_e B_0I_z,\label{H0}
	\end{equation}
	where ${\bf p}_e$ is the momentum, $m_e$ is the mass, $Z$ is the atomic number, $k$ is the Coulomb's constant, $e$ is the elementary charge, $\gamma_e$ is the gyromagnetic ratio, and $I_z$ is the orbital angular momentum. We do not consider electron spin since it does not couple with EF. The eigenvalue of this Hamiltonion is $E_{nm}=-\frac{Z^2m_ee^4}{2n^2\hbar^2}+\gamma_e B_0m\hbar$, and related eigenstate is $|nlm\rangle$. Here $n,l,m$ are principal, azimuthal and magnetic quantum number. Applying the external oscillating EF $(E_x(t),~E_y(t),~E_z(t))$ with angular frequency $\omega$, we get the perturbation Hamiltonion
	\begin{equation}
		\mathcal{H}_1=eE_\alpha(t) r_\alpha.
	\end{equation}
	The time evolution by the Hamiltonion $\mathcal{H}_0+\mathcal{H}_1$ is $|\psi(t)\rangle=\exp{\left\{-\frac{i}{\hbar}\int_0^{t}(\mathcal{H}_0+\mathcal{H}_1)dt\right\}}|\psi(0)\rangle$.
	The electron on the $E_{nm}$ energy level has $n-|m|$ degrees of degeneracy. Given initial state $|\psi(0)\rangle=\sum_{l=|m|}^{n-1}c_l|nlm\rangle$, we have:
	\begin{eqnarray}
		|\psi(t)\rangle=&&\exp{\left\{-\frac{i}{\hbar}E_{nm}t\right\}}\exp{\left\{\frac{ie}{\hbar\omega}\tilde{E}_\alpha(t)r_\alpha\right\}}\sum_lc_l|nlm\rangle\nonumber\\
		\approx&&\exp{(-\frac{i}{\hbar}E_{nm}t)}\sum_lc_l(1+\frac{ie}{\hbar\omega}\tilde{E}_\alpha(t)r_\alpha)|nlm\rangle,
	\end{eqnarray}
	where $\tilde{E}_\alpha(t)\equiv-\omega\int_0^{t}E_\alpha(t)dt$. After some simple calculations, we get EFG at the origin (the nuclear position) generated by the electron of state $|\psi(t)\rangle$:
	\begin{equation}
		\partial_\alpha E_\beta=ke\langle\psi(t)|\frac{3r_\alpha r_\beta}{r^5}-\frac{\delta_{\alpha\beta}}{r^3}|\psi(t)\rangle.\label{EFG_initial}
	\end{equation}
	Using $(r\cos\theta,~r\sin\theta\cos\phi,~r\sin\theta\sin\phi)$ coordinate and the recurrence formulas of associated Legendre polynomials we get each component of EFG to the first order after some tedious calculations:
	\begin{eqnarray}
		\partial_x E_x=&&\partial_y E_y=C(n,m)-B(n,m)\tilde{E}_z(t),\nonumber\\
		\partial_z E_z=&&-2C(n,m)+2B(n,m)\tilde{E}_z(t),\nonumber\\
		\partial_x E_y=&&\partial_y E_x=0,\partial_x E_z=\partial_z E_x=-3A(n,m)\tilde{E}_x(t),\nonumber\\
		\partial_y E_z=&&\partial_z E_y=-3A(n,m)\tilde{E}_y(t),
	\end{eqnarray}
	where
	\begin{eqnarray}
		A(n,m)\equiv&&\frac{ke^2}{\hbar\omega}\sum\limits_{l^\prime l}\langle nl^\prime m|\frac{\cos\theta-\cos^3\theta}{r^2}|nlm\rangle \mathrm{Im}(c^*_{l^\prime}c_l),\nonumber\\
		B(n,m)\equiv&&\frac{ke^2}{\hbar\omega}\sum\limits_{l^\prime l}\langle nl^\prime m|\frac{\cos\theta-3\cos^3\theta}{r^2}|nlm\rangle \mathrm{Im}(c^*_{l^\prime}c_l),\nonumber\\
		C(n,m)\equiv&&\frac{ke}{2}\sum_{l^\prime l}\langle nl^\prime m|\frac{1-3\cos^2\theta}{r^3}|nlm\rangle c^*_{l^\prime}c_l.\label{CAnm}
	\end{eqnarray}
	This is the EFG generated by an electron of state $|\psi(t)\rangle$. The total EFG is simply the summation for all electrons in the atom:
	\begin{eqnarray}
		\partial_x E_x=&&\partial_y E_y=C-B\tilde{E}_z(t),~\partial_z E_z=-2C+2B\tilde{E}_z(t),\nonumber\\
		\partial_x E_y=&&\partial_y E_x=0,~\partial_x E_z=\partial_z E_x=-3A\tilde{E}_x(t),\nonumber\\
		\partial_y E_z=&&\partial_z E_y=-3A\tilde{E}_y(t),\label{EFG_oscilating}
	\end{eqnarray}
	where
	\begin{equation}
		(A,~B,~C)=\sum_{\mathrm{all~ electrons}}(A(n,m),~B(n,m),~C(n,m)).\label{CA}
	\end{equation}
	With this total EFG, Eq.~(\ref{H_final}) leads to the Hamiltonion for nucleus:
	\begin{eqnarray}
		&&\mathcal{H}_n=\gamma_n B_0S_z+\tilde{Q}(C-B\tilde{E}_z(t))(-S_x^2-S_y^2+2S_z^2)\nonumber\\
		&&+3\tilde{Q}A[(S_xS_z+S_zS_x)\tilde{E}_x(t)+(S_yS_z+S_zS_y)\tilde{E}_y(t)]\nonumber\\
		&&~~~~=\gamma_n B_0S_z+3\tilde{Q}(C-B\tilde{E}_z(t))S_z^2+3\tilde{Q}A\times\nonumber\\
		&&[(S_xS_z+S_zS_x)\tilde{E}_x(t)+(S_yS_z+S_zS_y)\tilde{E}_y(t)].\label{H_NER_initial}
	\end{eqnarray}
	 In order to manipulate nuclear state, we take the external EF to be the following form:
	\begin{equation}
		(E_x(t),E_y(t),E_z(t))=E(\sin(\omega t+\varphi),-\cos(\omega t+\varphi),0).\label{EF_oscilating}
	\end{equation} 
	Then Eq.~(\ref{H_NER_initial}) becomes to
	\begin{eqnarray}
		&&\mathcal{H}_{\mathrm{NER}}=\gamma_n B_0S_z+3\tilde{Q}CS_z^2+3\tilde{Q}AE\times\nonumber\\
		&&[(S_xS_z+S_zS_x)\cos(\omega t+\varphi)+(S_yS_z+S_zS_y)\sin(\omega t+\varphi)],\nonumber\\\label{H_NER}
	\end{eqnarray}
	where the time-independent small quantities have been ignored. This is our major Hamiltonion for NER. Using the rotating frame $\Psi_{\mathrm{NER}}(t)=e^{-iS_z(\omega t)}\Psi'_{\mathrm{NER}}(t)$ in the Schr{\"{o}}dinger equation $i\hbar\partial_{t}\Psi_{\mathrm{NER}}(t)=\mathcal{H}_{\mathrm{NER}}\Psi_{\mathrm{NER}}(t)$ we obtain
	\begin{eqnarray}
		\Psi^\prime_{\mathrm{NER}}(t)=&&\exp\Big\{-\frac{it}{\hbar}3\tilde{Q}[CS_z^2+AE((S_xS_z+S_zS_x)\cos\varphi\nonumber\\
		&&+(S_yS_z+S_zS_y)\sin\varphi)]\Big\}\Psi'_{\mathrm{NER}}(0),
		\label{NER_function}
	\end{eqnarray}
	where resonance condition $\gamma_nB_0=\hbar\omega$ and the identity $I_x\cos(\omega t+\varphi)+I_y\sin(\omega t+\varphi)=e^{-iI_z(\omega t+\varphi)}I_xe^{iI_z(\omega t+\varphi)}$ have been used. With this we can precisely manipulate nuclear state. 
	
	Some notes about Eqs.~(\ref{H_NER}) and~(\ref{NER_function}):

	(i) The term $3\tilde{Q}CS_z^2$ is independent of our oscillating EF so it gives inhomogeneous energy level splittings all the time. This effect has been observed in experiment~\cite{asaad2020coherent}.
	
	(ii) The factor $3AE$ is the amplitude of EFG in Eq.~(\ref{EFG_oscilating}), which is much lager than the amplitude $E$ of oscillating EF in Eq.~(\ref{EF_oscilating}). We give a rough evaluation  $A\approx\frac{ke^2}{\hbar a_0^2\times10^7\mathrm{Hz}}\approx8\times10^{19}\mathrm{m}^{-1}$ where $a_0$ is the Bohr radius. This provides a sufficiently large Rabi frequency.
	
	(iii) NER will divide the Hilbert space into two parts if $S$ is a half-integer. Explicitly, the operator $(S_xS_z+S_zS_x)\cos\varphi+(S_yS_z+S_zS_y)\sin\varphi$ appearing in Eq.~(\ref{NER_function}) shows that the transitions can only appear between neighbor energy levels. But the matrix element of this operator will be zero if $m_S=-m^\prime_S=\frac{1}{2}$. So the subspaces of $m_S\le-\frac{1}{2}$ and of $m_S\ge\frac{1}{2}$ are independent. This agrees with the experiment~\cite{asaad2020coherent}. One more thing is that there does not exist NER for spin $S=\frac{1}{2}$.
	
	\section{IV. Nuclear energy splittings with external static electric field}
	
	Consider the following external static EF
	\begin{equation}
		(E_x(t),~E_y(t),~E_z(t))=(0,~0,~E_0).\label{static_EF}
	\end{equation}
	Now the perturbation Hamiltonion for an atomic electron changes to 
	\begin{equation}
		\mathcal{H}_1=ezE_0.
	\end{equation}
	The main Hamiltonion is still given by Eq.~(\ref{H0}). The degenerate subspace $\mathcal{D}$ of the $E_{nm}$ energy level has $n-|m|$ degrees of degeneracy. The $k$th energy and state in the space $\mathcal{D}$ are
	\begin{eqnarray}
		E_{nmk}=&&E^{(0)}+E^{(1)}_k,\nonumber\\
		|\psi_{nmk}\rangle=&&|\psi^{(0)}_{k}\rangle+|\psi^{(1)}_{k}\rangle,
	\end{eqnarray}
	where $E^{(0)}=E_{nm}$, $E^{(1)}_k$ and $|\psi^{(0)}_k\rangle$ are the $k$th eigenvalue and eigenstate for the projection matrix of $\mathcal{H}_1$ into space $\mathcal{D}$, i.e., $\mathcal{H}_1^{(\mathcal{D})}|\psi^{(0)}_k\rangle=E^{(1)}_k|\psi^{(0)}_k\rangle$. $|\psi^{(0)}_k\rangle$ can be written as $|\psi^{(0)}_k\rangle=\sum_lc_l|nlm\rangle$ and then
	\begin{eqnarray}
		|\psi^{(1)}_k\rangle=&&\sum_{n^\prime m^\prime k^\prime\notin \mathcal{D}}|\psi^{(0)}_{k^\prime}\rangle\frac{\langle\psi^{(0)}_{k^\prime}|\mathcal{H}_1|\psi^{(0)}_k\rangle}{E_{nm}-E_{n^\prime m^\prime}}\nonumber\\
		=&&eE_0\sum_{n^\prime\ne n}\sum_{l^\prime l}|n^\prime l^\prime m\rangle c_l\frac{\langle n^\prime l^\prime m|z|nlm\rangle}{E_{nm}-E_{n^\prime m}}.
	\end{eqnarray}
	Taking the same procedure as Eqs.~(\ref{EFG_initial})-(\ref{CA}), we obtain the total EFG on the nucleus:
	\begin{eqnarray}
		\partial_xE_x=&&\partial_yE_y=C+B^\prime E_0,~\partial_zE_z=-2(C+B^\prime E_0),\nonumber\\
		\partial_xE_y=&&\partial_yE_x=\partial_xE_z=\partial_zE_x=\partial_yE_z=\partial_zE_y=0,
	\end{eqnarray}
	where $C$ takes the same form as in Eqs.~(\ref{CAnm}) and~(\ref{CA}), and
	\begin{eqnarray}
		&&B^\prime\equiv ke^2\sum_{\mathrm{all~ electrons}}\sum_{n^\prime\ne n,l,l^\prime,l^{\prime\prime}}\nonumber\\
		&&\langle nl^{\prime\prime}m|\frac{1-3\cos^2\theta}{r^3}|n^\prime l^\prime m\rangle \frac{\langle n^\prime l^\prime m|z|nlm\rangle}{E_{nm}-E_{n^\prime m}}\mathrm{Re}(c^*_{l^{\prime\prime}}c_l).
	\end{eqnarray}
	Then the Hamiltonion for the nucleus in this atom is:
	\begin{equation}
		\mathcal{H}_n=\gamma_n B_0S_z+3\tilde{Q}CS_z^2+3\tilde{Q}B^\prime E_0S_z^2.\label{H_LQSE}
	\end{equation}
	Remark:
	
	(i) The term $3\tilde{Q}CS_z^2$ is independent of our static EF. It is a natural effect for a nucleus in an atom. This agrees with Eq.~(\ref{H_NER}).
	
	(ii) The phenomenon that static EFG induces inhomogeneous nuclear energy level splittings is called linear quadrupole stark effect (LQSE)~\cite{Slichter2013}. However, with the term $3\tilde{Q}B^\prime E_0S_z^2$ appearing in above Hamiltonion we can control these nuclear energy level splittings by a simple EF instead of an unachievable EFG. A rough evaluation shows that $B^\prime\approx\frac{ke^2}{(E_{2m}-E_{1m}) a_0^2}\approx10^{20}\mathrm{m}^{-1}$ makes a remarkable energy gap.
	
	These two terms are useful in quantum computing as shown in section V and VI. 
	
	\section{V. Single-qubit Operation for $S\ge1$}
	
	Combining Eqs.~(\ref{EF_oscilating}) and~(\ref{static_EF}) we obtain the external electric field:
	\begin{eqnarray}
		E_x(t)=&&E\sin(\omega t+\varphi),~E_z(t)=E_0,\nonumber\\
		E_y(t)=&&-E\cos(\omega t+\varphi).\label{single_EF}
	\end{eqnarray}
	Given Eqs.~(\ref{H_NER}) and~(\ref{H_LQSE}) we have the following Hamiltonion for nucleus:
	\begin{eqnarray}
		&&\mathcal{H}_{\mathrm{single}}=\gamma_n B_0S_z+3\tilde{Q}(C+B^\prime E_0)S_z^2+3\tilde{Q}AE\times\nonumber\\
		&&[(S_xS_z+S_zS_x)\cos(\omega t+\varphi)+(S_yS_z+S_zS_y)\sin(\omega t+\varphi)].\nonumber\\\label{H_single}
	\end{eqnarray}
	The term $\gamma_n B_0S_z+3\tilde{Q}(C+B^\prime E_0)S_z^2$ makes the energy level splitting between $m_S$ and $m_S+1$ to be $\gamma_nB_0\hbar+3\tilde{Q}(C+B^\prime E_0)(2m_S+1)\hbar^2$. So we select two specific energy levels by applying specific frequency $\omega$. We choose the subspace $\mathcal{O}$ of $m_S=\{S,S-1\}$ for our qubit. The projection matrices of $S_x,~S_y,~S_z$ into this subspace are:
	\begin{eqnarray}
		S_x^{(\mathcal{O})}=&&\sqrt{2S}S_x^{(\frac{1}{2})},~~S_y^{(\mathcal{O})}=\sqrt{2S}S_y^{(\frac{1}{2})},\nonumber\\
		S_z^{(\mathcal{O})}=&&S_z^{(\frac{1}{2})}+\frac{\hbar}{2}(2S-1),
		\label{subspace_angular_momumtum}
	\end{eqnarray}
	where $S_x^{(\frac{1}{2})},~S_y^{(\frac{1}{2})},~S_z^{(\frac{1}{2})}$ are the spin $\frac{1}{2}$ matrices. In the $\mathcal{O}$ space Eq.~(\ref{H_single}) becomes to
	\begin{eqnarray}
		\mathcal{H}_{\mathrm{single}}^{(\mathcal{O})}=&&\gamma_n B_0S_z^{(\mathcal{O})}+3\tilde{Q}(C+B^\prime E_0)S_z^{(\mathcal{O})2}\nonumber\\
		&&+3\tilde{Q}AE[(S_x^{(\mathcal{O})}S_z^{(\mathcal{O})}+S_z^{(\mathcal{O})}S_x^{(\mathcal{O})})\cos(\omega t+\varphi)\nonumber\\
		&&+(S_y^{(\mathcal{O})}S_z^{(\mathcal{O})}+S_z^{(\mathcal{O})}S_y^{(\mathcal{O})})\sin(\omega t+\varphi)]\nonumber\\
		=&&[\gamma_n B_0+3(2S-1)\hbar\tilde{Q}(C+B^\prime E_0)]S_z^{(\frac{1}{2})}\nonumber\\
		&&+3\sqrt{2S}(2S-1)\hbar\tilde{Q}AE\times\nonumber\\
		&&\left[S_x^{(\frac{1}{2})}\cos(\omega t+\varphi)+S_y^{(\frac{1}{2})}\sin(\omega t+\varphi)\right].
		\label{H_single_O}
	\end{eqnarray}
	Using the rotating frame $\Psi_{\mathrm{single}}(t)=e^{-iS_z^{(\mathcal{O})}(\omega t)}\Psi^{(\mathcal{O})\prime}_{\mathrm{single}}(t)$ in the Schr{\"{o}}dinger equation $i\hbar\partial_t\Psi_{\mathrm{single}}^{(\mathcal{O})}(t)=\mathcal{H}_{\mathrm{single}}^{(\mathcal{O})}\Psi_{\mathrm{single}}^{(\mathcal{O})}(t)$ we obtain
	\begin{eqnarray}
		\Psi_{\mathrm{single}}^{(\mathcal{O})\prime}(t)=&&\exp\Big\{-\frac{it}{\hbar}3\sqrt{2S}(2S-1)\hbar\tilde{Q}AE\times\nonumber\\
		&&\left(S_x^{(\frac{1}{2})}\cos\varphi+S_y^{(\frac{1}{2})}\sin\varphi\right)\Big\}\Psi_{\mathrm{single}}^{(\mathcal{O})\prime}(0),
		\label{single_function}
	\end{eqnarray}
	in the subspace $\mathcal{O}$. Here resonance condition $\gamma_n B_0+3(2S-1)\hbar\tilde{Q}(C+B^\prime E_0)=\hbar\omega$ has been used. With Eq.~(\ref{single_function}), any single-qubit gate can be achieved for nuclear spin $S\ge1$.
	
	There is a more remarkable result for $S=\frac{3}{2}$. We apply appropriate frequency EF to select the subspace $\mathcal{O}:\{m_S=\frac{3}{2},~\frac{1}{2}\}$ and neglect the other subspace $\mathcal{O}^\prime:\{m_S=-\frac{1}{2},~-\frac{3}{2}\}$. But as we mentioned earlier in section III, NER naturally divides the Hilbert space of $S=\frac{3}{2}$ into two unconnected subspaces. So subspace $\mathcal{O}$ will be protected most robustly if $S=\frac{3}{2}$.
	
	\section{VI. Two-qubit Operation for $S\ge1$}
	
	Now we show two-qubit operations in $\mathcal{O}_1\otimes\mathcal{O}_2$ of nucleus $1$ and nucleus $2$, which is not disturbed by other subspaces. This proposal can be realized in solid-state system~\cite{asaad2020coherent}. We apply the EF $(0,0,E_1)$ and $(0,0,E_2)$ to two nuclei separately. Using Eq.~(\ref{H_LQSE}) we obtain the Hamiltonion for two nuclei with interaction:
	\begin{eqnarray}
		\mathcal{H}_{\mathrm{two}}=&&\gamma_{n1}B_0S_{1z}+\gamma_{n2}B_0S_{2z}+3\tilde{Q}_1C_1S_{1z}^2+3\tilde{Q}_2C_2S_{2z}^2\nonumber\\
		&&+3\tilde{Q}_1B^\prime_1 E_1S_{1z}^2+3\tilde{Q}_2B^\prime_2E_2S_{2z}^2+\mathcal{H}_\mathrm{J},
		\label{H_two}
	\end{eqnarray}
	where index $1$ and $2$ represent different nuclei, $\mathcal{H}_\mathrm{J}$ is the J-coupling interaction. We use two different kinds of nuclei or apply much different $E_1$ and $E_2$ to same kinds of nuclei. This means that the difference between two nuclear energy level splittings is much larger than J-coupling interaction and hence we can take the approximation $\mathcal{H}_\mathrm{J}\approx 2\pi J(t)S_{1z}S_{2z}$. Notably, $\mathcal{H}_\mathrm{J}$ decouples the subspace $\mathcal{O}_1\otimes\mathcal{O}_2$ from other subspaces since it is a diagonal matrix. Similar to Eq.~(\ref{H_single_O}), we consider Eq.~(\ref{H_two}) in $\mathcal{O}_1\otimes\mathcal{O}_2$:
	\begin{eqnarray}
		\mathcal{H}_{\mathrm{two}}^{(\mathcal{O}_1\otimes\mathcal{O}_2)}=&&[\gamma_{n1}B_0+3(2S-1)\hbar\tilde{Q}_1C_1]\left(S_{1z}^{(\frac{1}{2})}+S_{2z}^{(\frac{1}{2})}\right)\nonumber\\
		+[(\gamma_{n2}&&-\gamma_{n1})B_0+3(2S-1)\hbar(\tilde{Q}_2C_2-\tilde{Q}_1C_1)]S_{2z}^{(\frac{1}{2})}\nonumber\\
		+&&(2S-1)\hbar(3\tilde{Q}_1B^\prime_1 E_1+\pi J(t))S_{1z}^{(\frac{1}{2})}\nonumber\\
		+&&(2S-1)\hbar(3\tilde{Q}_2B^\prime_2 E_2+\pi J(t))S_{2z}^{(\frac{1}{2})}\nonumber\\
		+&&2\pi J(t)S_{1z}^{(\frac{1}{2})}S_{2z}^{(\frac{1}{2})},
		\label{H_J_2}
	\end{eqnarray}
	where Eq.~(\ref{subspace_angular_momumtum}) has been used. Using the rotating frame $\Psi_{\mathrm{two}}^{(\mathcal{O}_1\otimes\mathcal{O}_2)}(t)=e^{-i\left(S_{1z}^{(\frac{1}{2})}+S_{2z}^{(\frac{1}{2})}\right)(\omega t)}\Psi^{(\mathcal{O}_1\otimes\mathcal{O}_2)\prime}_{\mathrm{two}}(t)$ in the Schr{\"{o}}dinger equation
	$i\hbar\partial_t\Psi_{\mathrm{two}}^{(\mathcal{O}_1\otimes\mathcal{O}_2)}(t)=\mathcal{H}_{\mathrm{two}}\Psi_{\mathrm{two}}^{(\mathcal{O}_1\otimes\mathcal{O}_2)}(t)$ we obtain
	\begin{equation}
		\Psi^{(\mathcal{O}_1\otimes\mathcal{O}_2)\prime}_{\mathrm{two}}(t)=U_1(t)U_2(t)U_{1,2}(t)\Psi^{(\mathcal{O}_1\otimes\mathcal{O}_2)\prime}_{\mathrm{two}}(0),
		\label{solution_J}
	\end{equation}
	where
	\begin{equation}
		U_1(t)=\exp\Big\{-\frac{i}{\hbar}S_{1z}^{(\frac{1}{2})}\int^t_0(2S-1)\hbar
		(3\tilde{Q}_1B^\prime_1 E_1+\pi J(t))dt\Big\}
	\end{equation}
	is the operation for the first qubit,
	\begin{eqnarray}
		U_2(t)=&&\exp\Big\{-\frac{i}{\hbar}S_{2z}^{(\frac{1}{2})}\int^t_0(\gamma_{n2}-\gamma_{n1})B_0+(2S-1)\hbar\times\nonumber\\
		&&[3\tilde{Q}_2B^\prime_2 E_2+\pi J(t)+3(\tilde{Q}_2C_2-\tilde{Q}_1C_1)]dt\Big\}
	\end{eqnarray}
		is the operation for the second qubit, and
	\begin{equation}
		 U_{1,2}(t)=\exp\left\{-\frac{i}{\hbar}S_{1z}^{(\frac{1}{2})}S_{2z}^{(\frac{1}{2})}\int^t_02\pi J(t)dt\right\}
	\end{equation}
	is the two-qubit operation. Here resonance condition $\gamma_{n1}B_0+(2S-1)\hbar\tilde{Q}_1E_0=\hbar\omega$ has been used. By tuning $E_1,~E_2,~J(t)$ we can get a controlled-NOT gate or a controlled-Z gate~\cite{Nielsen2002,loss1998quantum}.
	
	\section{VII. Performance Evaluation}
	
	In quantum computing, all operations should be completed within the coherence time $T_2^*$. We can calculate the number of flips $N_\mathrm{f}\equiv T_2^*\times f_\mathrm{R}$ of our NER proposal, where $f_\mathrm{R}$ is the Rabi frequency. It is the number of Rabi oscillations during a coherence time $T_2^*$. After this calculation, we compare NER with ENMHSE as shown below.
	
	From the study on Sb nucleus in Ref.~\cite{asaad2020coherent}, we get the following data: $f_\mathrm{R}=684.2$Hz, $T_2^*=92$ms under the condition $V_\mathrm{RF}=20$mV which is the voltage of radio-frequency (RF) oscillating EF. According to Eq.~(\ref{single_function}) we obtain $f_\mathrm{R}=k_RE$ where $k_R\equiv\left|\frac{3eQA}{2\pi\sqrt{2S}\hbar}\right|$. To our knowledge the voltage of oscillating EF can easily reach $V^\prime_\mathrm{RF}=4\mathrm{V}$~\cite{sigillito2017all}, and the amplitude of EF is changed to $E^\prime=\frac{V^\prime_\mathrm{RF}}{V_\mathrm{RF}}E=200E$. The Rabi frequency is improved to $f^\prime_\mathrm{R}=k_RE^\prime=200k_RE=200f_\mathrm{R}=136840\mathrm{Hz}$. Finally we calculate $N_\mathrm{f}=T_2^*\times f^\prime_\mathrm{R}=12589.28$ for our NER proposal. Comparing this result with ENMHSE as shown in Table~\ref{tab:1}, we find an improvement by three orders of magnitude on $N_\mathrm{f}$.
	\begin{table}[!]
	\caption{\label{tab:1}%
		Number of flips with coherence time $T_2^*$: comparison of our NER proposal with existing ENMHSE~\cite{thiele2014electrically,laucht2015electrically}. The nuclei are Tb, P and Sb respectively.}
		\begin{ruledtabular}
			\begin{tabular}{llll}
				Method &
				Coherence time &
				Rabi frequency &
				Number of flips\\
				& $T_2^*/\mathrm{ms}$ & $f_\mathrm{R}/\mathrm{kHz}$ & $N_\mathrm{f}=T_2^*\times f_\mathrm{R}$\\
				\colrule
				ENMHSE &$\sim0.064$ &$\sim180.8$  &$\sim11.57$\\
				ENMHSE &$\sim0.97$ &$\sim5$ &$\sim4.85$\\
				NER &$\sim92$ &$\sim136.8$ &$\sim12589.28$\\
			\end{tabular}
		\end{ruledtabular}
	\end{table}
	
	\section{VIII. Conclusion}
	
	In summary, we present a new nuclear manipulation proposal called NER, and apply it in constructing universal quantum gates on arbitrary $S\ge 1$ spin nuclei. Since NER can manipulate single nucleus and meanwhile hold a long coherence time, it is quiet useful for quantum information processing in the future.
	
	\begin{acknowledgements}
		We acknowledge the financial support in part by Ministry of Science and Technology of China through The National Key Research and Development Program of China grant No. 2017YFA0303901; National Natural Science Foundation of China grant No.11774198 and No.11974204.
	\end{acknowledgements}

	\bibliography{refs}

\begin{thebibliography}{18}%
\makeatletter
\providecommand \@ifxundefined [1]{%
 \@ifx{#1\undefined}
}%
\providecommand \@ifnum [1]{%
 \ifnum #1\expandafter \@firstoftwo
 \else \expandafter \@secondoftwo
 \fi
}%
\providecommand \@ifx [1]{%
 \ifx #1\expandafter \@firstoftwo
 \else \expandafter \@secondoftwo
 \fi
}%
\providecommand \natexlab [1]{#1}%
\providecommand \enquote  [1]{``#1''}%
\providecommand \bibnamefont  [1]{#1}%
\providecommand \bibfnamefont [1]{#1}%
\providecommand \citenamefont [1]{#1}%
\providecommand \href@noop [0]{\@secondoftwo}%
\providecommand \href [0]{\begingroup \@sanitize@url \@href}%
\providecommand \@href[1]{\@@startlink{#1}\@@href}%
\providecommand \@@href[1]{\endgroup#1\@@endlink}%
\providecommand \@sanitize@url [0]{\catcode `\\12\catcode `\$12\catcode
  `\&12\catcode `\#12\catcode `\^12\catcode `\_12\catcode `\%12\relax}%
\providecommand \@@startlink[1]{}%
\providecommand \@@endlink[0]{}%
\providecommand \url  [0]{\begingroup\@sanitize@url \@url }%
\providecommand \@url [1]{\endgroup\@href {#1}{\urlprefix }}%
\providecommand \urlprefix  [0]{URL }%
\providecommand \Eprint [0]{\href }%
\providecommand \doibase [0]{https://doi.org/}%
\providecommand \selectlanguage [0]{\@gobble}%
\providecommand \bibinfo  [0]{\@secondoftwo}%
\providecommand \bibfield  [0]{\@secondoftwo}%
\providecommand \translation [1]{[#1]}%
\providecommand \BibitemOpen [0]{}%
\providecommand \bibitemStop [0]{}%
\providecommand \bibitemNoStop [0]{.\EOS\space}%
\providecommand \EOS [0]{\spacefactor3000\relax}%
\providecommand \BibitemShut  [1]{\csname bibitem#1\endcsname}%
\let\auto@bib@innerbib\@empty
\bibitem [{\citenamefont {DiVincenzo}(1995)}]{divincenzo1995quantum}%
  \BibitemOpen
  \bibfield  {author} {\bibinfo {author} {\bibfnamefont {D.~P.}\ \bibnamefont
  {DiVincenzo}},\ }\href@noop {} {\bibfield  {journal} {\bibinfo  {journal}
  {Science}\ }\textbf {\bibinfo {volume} {270}},\ \bibinfo {pages} {255}
  (\bibinfo {year} {1995})}\BibitemShut {NoStop}%
\bibitem [{\citenamefont {Ekert}\ and\ \citenamefont
  {Jozsa}(1996)}]{ekert1996}%
  \BibitemOpen
  \bibfield  {author} {\bibinfo {author} {\bibfnamefont {A.}~\bibnamefont
  {Ekert}}\ and\ \bibinfo {author} {\bibfnamefont {R.}~\bibnamefont {Jozsa}},\
  }\href@noop {} {\bibfield  {journal} {\bibinfo  {journal} {Reviews of Modern
  Physics}\ }\textbf {\bibinfo {volume} {68}},\ \bibinfo {pages} {733}
  (\bibinfo {year} {1996})}\BibitemShut {NoStop}%
\bibitem [{\citenamefont {Preskill}(2018)}]{preskill2018}%
  \BibitemOpen
  \bibfield  {author} {\bibinfo {author} {\bibfnamefont {J.}~\bibnamefont
  {Preskill}},\ }\href@noop {} {\bibfield  {journal} {\bibinfo  {journal}
  {Quantum}\ }\textbf {\bibinfo {volume} {2}},\ \bibinfo {pages} {79} (\bibinfo
  {year} {2018})}\BibitemShut {NoStop}%
\bibitem [{\citenamefont {Arute}\ \emph {et~al.}(2019)\citenamefont {Arute},
  \citenamefont {Arya}, \citenamefont {Babbush}, \citenamefont {Bacon},
  \citenamefont {Bardin}, \citenamefont {Barends}, \citenamefont {Biswas},
  \citenamefont {Boixo}, \citenamefont {Brandao}, \citenamefont {Buell} \emph
  {et~al.}}]{arute2019quantum}%
  \BibitemOpen
  \bibfield  {author} {\bibinfo {author} {\bibfnamefont {F.}~\bibnamefont
  {Arute}}, \bibinfo {author} {\bibfnamefont {K.}~\bibnamefont {Arya}},
  \bibinfo {author} {\bibfnamefont {R.}~\bibnamefont {Babbush}}, \bibinfo
  {author} {\bibfnamefont {D.}~\bibnamefont {Bacon}}, \bibinfo {author}
  {\bibfnamefont {J.~C.}\ \bibnamefont {Bardin}}, \bibinfo {author}
  {\bibfnamefont {R.}~\bibnamefont {Barends}}, \bibinfo {author} {\bibfnamefont
  {R.}~\bibnamefont {Biswas}}, \bibinfo {author} {\bibfnamefont
  {S.}~\bibnamefont {Boixo}}, \bibinfo {author} {\bibfnamefont {F.~G.}\
  \bibnamefont {Brandao}}, \bibinfo {author} {\bibfnamefont {D.~A.}\
  \bibnamefont {Buell}}, \emph {et~al.},\ }\href@noop {} {\bibfield  {journal}
  {\bibinfo  {journal} {Nature}\ }\textbf {\bibinfo {volume} {574}},\ \bibinfo
  {pages} {505} (\bibinfo {year} {2019})}\BibitemShut {NoStop}%
\bibitem [{\citenamefont {Wu}\ \emph {et~al.}(2021)\citenamefont {Wu},
  \citenamefont {Bao}, \citenamefont {Cao}, \citenamefont {Chen}, \citenamefont
  {Chen}, \citenamefont {Chen}, \citenamefont {Chung}, \citenamefont {Deng},
  \citenamefont {Du}, \citenamefont {Fan} \emph {et~al.}}]{wu2021strong}%
  \BibitemOpen
  \bibfield  {author} {\bibinfo {author} {\bibfnamefont {Y.}~\bibnamefont
  {Wu}}, \bibinfo {author} {\bibfnamefont {W.-S.}\ \bibnamefont {Bao}},
  \bibinfo {author} {\bibfnamefont {S.}~\bibnamefont {Cao}}, \bibinfo {author}
  {\bibfnamefont {F.}~\bibnamefont {Chen}}, \bibinfo {author} {\bibfnamefont
  {M.-C.}\ \bibnamefont {Chen}}, \bibinfo {author} {\bibfnamefont
  {X.}~\bibnamefont {Chen}}, \bibinfo {author} {\bibfnamefont {T.-H.}\
  \bibnamefont {Chung}}, \bibinfo {author} {\bibfnamefont {H.}~\bibnamefont
  {Deng}}, \bibinfo {author} {\bibfnamefont {Y.}~\bibnamefont {Du}}, \bibinfo
  {author} {\bibfnamefont {D.}~\bibnamefont {Fan}}, \emph {et~al.},\
  }\href@noop {} {\bibfield  {journal} {\bibinfo  {journal} {Physical Review
  Letters}\ }\textbf {\bibinfo {volume} {127}},\ \bibinfo {pages} {180501}
  (\bibinfo {year} {2021})}\BibitemShut {NoStop}%
\bibitem [{\citenamefont {Zhong}\ \emph {et~al.}(2021)\citenamefont {Zhong},
  \citenamefont {Deng}, \citenamefont {Qin}, \citenamefont {Wang},
  \citenamefont {Chen}, \citenamefont {Peng}, \citenamefont {Luo},
  \citenamefont {Wu}, \citenamefont {Gong}, \citenamefont {Su} \emph
  {et~al.}}]{zhong2021phase}%
  \BibitemOpen
  \bibfield  {author} {\bibinfo {author} {\bibfnamefont {H.-S.}\ \bibnamefont
  {Zhong}}, \bibinfo {author} {\bibfnamefont {Y.-H.}\ \bibnamefont {Deng}},
  \bibinfo {author} {\bibfnamefont {J.}~\bibnamefont {Qin}}, \bibinfo {author}
  {\bibfnamefont {H.}~\bibnamefont {Wang}}, \bibinfo {author} {\bibfnamefont
  {M.-C.}\ \bibnamefont {Chen}}, \bibinfo {author} {\bibfnamefont {L.-C.}\
  \bibnamefont {Peng}}, \bibinfo {author} {\bibfnamefont {Y.-H.}\ \bibnamefont
  {Luo}}, \bibinfo {author} {\bibfnamefont {D.}~\bibnamefont {Wu}}, \bibinfo
  {author} {\bibfnamefont {S.-Q.}\ \bibnamefont {Gong}}, \bibinfo {author}
  {\bibfnamefont {H.}~\bibnamefont {Su}}, \emph {et~al.},\ }\href@noop {}
  {\bibfield  {journal} {\bibinfo  {journal} {Physical Review Letters}\
  }\textbf {\bibinfo {volume} {127}},\ \bibinfo {pages} {180502} (\bibinfo
  {year} {2021})}\BibitemShut {NoStop}%
\bibitem [{\citenamefont {Nielsen}\ and\ \citenamefont
  {Chuang}(2002)}]{Nielsen2002}%
  \BibitemOpen
  \bibfield  {author} {\bibinfo {author} {\bibfnamefont {M.~A.}\ \bibnamefont
  {Nielsen}}\ and\ \bibinfo {author} {\bibfnamefont {I.}~\bibnamefont
  {Chuang}},\ }\href@noop {} {\emph {\bibinfo {title} {Quantum computation and
  quantum information}}}\ (\bibinfo  {publisher} {American Association of
  Physics Teachers},\ \bibinfo {year} {2002})\BibitemShut {NoStop}%
\bibitem [{\citenamefont {Park}\ \emph {et~al.}(2017)\citenamefont {Park},
  \citenamefont {Yan}, \citenamefont {Loh}, \citenamefont {Will},\ and\
  \citenamefont {Zwierlein}}]{park2017second}%
  \BibitemOpen
  \bibfield  {author} {\bibinfo {author} {\bibfnamefont {J.~W.}\ \bibnamefont
  {Park}}, \bibinfo {author} {\bibfnamefont {Z.~Z.}\ \bibnamefont {Yan}},
  \bibinfo {author} {\bibfnamefont {H.}~\bibnamefont {Loh}}, \bibinfo {author}
  {\bibfnamefont {S.~A.}\ \bibnamefont {Will}},\ and\ \bibinfo {author}
  {\bibfnamefont {M.~W.}\ \bibnamefont {Zwierlein}},\ }\href@noop {} {\bibfield
   {journal} {\bibinfo  {journal} {Science}\ }\textbf {\bibinfo {volume}
  {357}},\ \bibinfo {pages} {372} (\bibinfo {year} {2017})}\BibitemShut
  {NoStop}%
\bibitem [{\citenamefont {Sharma}\ \emph {et~al.}(2019)\citenamefont {Sharma},
  \citenamefont {Gaebel}, \citenamefont {Rej}, \citenamefont {Reilly},
  \citenamefont {Economou},\ and\ \citenamefont
  {Barnes}}]{sharma2019enhancement}%
  \BibitemOpen
  \bibfield  {author} {\bibinfo {author} {\bibfnamefont {G.}~\bibnamefont
  {Sharma}}, \bibinfo {author} {\bibfnamefont {T.}~\bibnamefont {Gaebel}},
  \bibinfo {author} {\bibfnamefont {E.}~\bibnamefont {Rej}}, \bibinfo {author}
  {\bibfnamefont {D.~J.}\ \bibnamefont {Reilly}}, \bibinfo {author}
  {\bibfnamefont {S.~E.}\ \bibnamefont {Economou}},\ and\ \bibinfo {author}
  {\bibfnamefont {E.}~\bibnamefont {Barnes}},\ }\href@noop {} {\bibfield
  {journal} {\bibinfo  {journal} {Physical Review B}\ }\textbf {\bibinfo
  {volume} {99}},\ \bibinfo {pages} {205423} (\bibinfo {year}
  {2019})}\BibitemShut {NoStop}%
\bibitem [{\citenamefont {Serafin}\ \emph {et~al.}(2021)\citenamefont
  {Serafin}, \citenamefont {Fadel}, \citenamefont {Treutlein},\ and\
  \citenamefont {Sinatra}}]{serafin2021nuclear}%
  \BibitemOpen
  \bibfield  {author} {\bibinfo {author} {\bibfnamefont {A.}~\bibnamefont
  {Serafin}}, \bibinfo {author} {\bibfnamefont {M.}~\bibnamefont {Fadel}},
  \bibinfo {author} {\bibfnamefont {P.}~\bibnamefont {Treutlein}},\ and\
  \bibinfo {author} {\bibfnamefont {A.}~\bibnamefont {Sinatra}},\ }\href@noop
  {} {\bibfield  {journal} {\bibinfo  {journal} {Physical Review Letters}\
  }\textbf {\bibinfo {volume} {127}},\ \bibinfo {pages} {013601} (\bibinfo
  {year} {2021})}\BibitemShut {NoStop}%
\bibitem [{\citenamefont {Thiele}\ \emph {et~al.}(2014)\citenamefont {Thiele},
  \citenamefont {Balestro}, \citenamefont {Ballou}, \citenamefont {Klyatskaya},
  \citenamefont {Ruben},\ and\ \citenamefont
  {Wernsdorfer}}]{thiele2014electrically}%
  \BibitemOpen
  \bibfield  {author} {\bibinfo {author} {\bibfnamefont {S.}~\bibnamefont
  {Thiele}}, \bibinfo {author} {\bibfnamefont {F.}~\bibnamefont {Balestro}},
  \bibinfo {author} {\bibfnamefont {R.}~\bibnamefont {Ballou}}, \bibinfo
  {author} {\bibfnamefont {S.}~\bibnamefont {Klyatskaya}}, \bibinfo {author}
  {\bibfnamefont {M.}~\bibnamefont {Ruben}},\ and\ \bibinfo {author}
  {\bibfnamefont {W.}~\bibnamefont {Wernsdorfer}},\ }\href@noop {} {\bibfield
  {journal} {\bibinfo  {journal} {Science}\ }\textbf {\bibinfo {volume}
  {344}},\ \bibinfo {pages} {1135} (\bibinfo {year} {2014})}\BibitemShut
  {NoStop}%
\bibitem [{\citenamefont {Laucht}\ \emph {et~al.}(2015)\citenamefont {Laucht},
  \citenamefont {Muhonen}, \citenamefont {Mohiyaddin}, \citenamefont {Kalra},
  \citenamefont {Dehollain}, \citenamefont {Freer}, \citenamefont {Hudson},
  \citenamefont {Veldhorst}, \citenamefont {Rahman}, \citenamefont {Klimeck}
  \emph {et~al.}}]{laucht2015electrically}%
  \BibitemOpen
  \bibfield  {author} {\bibinfo {author} {\bibfnamefont {A.}~\bibnamefont
  {Laucht}}, \bibinfo {author} {\bibfnamefont {J.~T.}\ \bibnamefont {Muhonen}},
  \bibinfo {author} {\bibfnamefont {F.~A.}\ \bibnamefont {Mohiyaddin}},
  \bibinfo {author} {\bibfnamefont {R.}~\bibnamefont {Kalra}}, \bibinfo
  {author} {\bibfnamefont {J.~P.}\ \bibnamefont {Dehollain}}, \bibinfo {author}
  {\bibfnamefont {S.}~\bibnamefont {Freer}}, \bibinfo {author} {\bibfnamefont
  {F.~E.}\ \bibnamefont {Hudson}}, \bibinfo {author} {\bibfnamefont
  {M.}~\bibnamefont {Veldhorst}}, \bibinfo {author} {\bibfnamefont
  {R.}~\bibnamefont {Rahman}}, \bibinfo {author} {\bibfnamefont
  {G.}~\bibnamefont {Klimeck}}, \emph {et~al.},\ }\href@noop {} {\bibfield
  {journal} {\bibinfo  {journal} {Science Advances}\ }\textbf {\bibinfo
  {volume} {1}},\ \bibinfo {pages} {e1500022} (\bibinfo {year}
  {2015})}\BibitemShut {NoStop}%
\bibitem [{\citenamefont {Sigillito}\ \emph {et~al.}(2017)\citenamefont
  {Sigillito}, \citenamefont {Tyryshkin}, \citenamefont {Schenkel},
  \citenamefont {Houck},\ and\ \citenamefont {Lyon}}]{sigillito2017all}%
  \BibitemOpen
  \bibfield  {author} {\bibinfo {author} {\bibfnamefont {A.~J.}\ \bibnamefont
  {Sigillito}}, \bibinfo {author} {\bibfnamefont {A.~M.}\ \bibnamefont
  {Tyryshkin}}, \bibinfo {author} {\bibfnamefont {T.}~\bibnamefont {Schenkel}},
  \bibinfo {author} {\bibfnamefont {A.~A.}\ \bibnamefont {Houck}},\ and\
  \bibinfo {author} {\bibfnamefont {S.~A.}\ \bibnamefont {Lyon}},\ }\href@noop
  {} {\bibfield  {journal} {\bibinfo  {journal} {Nature Nanotechnology}\
  }\textbf {\bibinfo {volume} {12}},\ \bibinfo {pages} {958} (\bibinfo {year}
  {2017})}\BibitemShut {NoStop}%
\bibitem [{\citenamefont {Asaad}\ \emph {et~al.}(2020)\citenamefont {Asaad},
  \citenamefont {Mourik}, \citenamefont {Joecker}, \citenamefont {Johnson},
  \citenamefont {Baczewski}, \citenamefont {Firgau}, \citenamefont
  {M{\k{a}}dzik}, \citenamefont {Schmitt}, \citenamefont {Pla}, \citenamefont
  {Hudson} \emph {et~al.}}]{asaad2020coherent}%
  \BibitemOpen
  \bibfield  {author} {\bibinfo {author} {\bibfnamefont {S.}~\bibnamefont
  {Asaad}}, \bibinfo {author} {\bibfnamefont {V.}~\bibnamefont {Mourik}},
  \bibinfo {author} {\bibfnamefont {B.}~\bibnamefont {Joecker}}, \bibinfo
  {author} {\bibfnamefont {M.~A.}\ \bibnamefont {Johnson}}, \bibinfo {author}
  {\bibfnamefont {A.~D.}\ \bibnamefont {Baczewski}}, \bibinfo {author}
  {\bibfnamefont {H.~R.}\ \bibnamefont {Firgau}}, \bibinfo {author}
  {\bibfnamefont {M.~T.}\ \bibnamefont {M{\k{a}}dzik}}, \bibinfo {author}
  {\bibfnamefont {V.}~\bibnamefont {Schmitt}}, \bibinfo {author} {\bibfnamefont
  {J.~J.}\ \bibnamefont {Pla}}, \bibinfo {author} {\bibfnamefont {F.~E.}\
  \bibnamefont {Hudson}}, \emph {et~al.},\ }\href@noop {} {\bibfield  {journal}
  {\bibinfo  {journal} {Nature}\ }\textbf {\bibinfo {volume} {579}},\ \bibinfo
  {pages} {205} (\bibinfo {year} {2020})}\BibitemShut {NoStop}%
\bibitem [{\citenamefont {Bloembergen}(1961)}]{Bloembergen1961}%
  \BibitemOpen
  \bibfield  {author} {\bibinfo {author} {\bibfnamefont {N.}~\bibnamefont
  {Bloembergen}},\ }\href@noop {} {\bibfield  {journal} {\bibinfo  {journal}
  {Science}\ }\textbf {\bibinfo {volume} {133}},\ \bibinfo {pages} {1363}
  (\bibinfo {year} {1961})}\BibitemShut {NoStop}%
\bibitem [{\citenamefont {Slichter}(2013)}]{Slichter2013}%
  \BibitemOpen
  \bibfield  {author} {\bibinfo {author} {\bibfnamefont {C.~P.}\ \bibnamefont
  {Slichter}},\ }\href@noop {} {\emph {\bibinfo {title} {Principles of magnetic
  resonance}}},\ Vol.~\bibinfo {volume} {1}\ (\bibinfo  {publisher} {Springer
  Science \& Business Media},\ \bibinfo {year} {2013})\BibitemShut {NoStop}%
\bibitem [{\citenamefont {Scully}\ and\ \citenamefont
  {Zubairy}(1999)}]{Scully1997}%
  \BibitemOpen
  \bibfield  {author} {\bibinfo {author} {\bibfnamefont {M.~O.}\ \bibnamefont
  {Scully}}\ and\ \bibinfo {author} {\bibfnamefont {M.~S.}\ \bibnamefont
  {Zubairy}},\ }\href@noop {} {\emph {\bibinfo {title} {Quantum optics}}}\
  (\bibinfo  {publisher} {American Association of Physics Teachers},\ \bibinfo
  {year} {1999})\BibitemShut {NoStop}%
\bibitem [{\citenamefont {Loss}\ and\ \citenamefont
  {DiVincenzo}(1998)}]{loss1998quantum}%
  \BibitemOpen
  \bibfield  {author} {\bibinfo {author} {\bibfnamefont {D.}~\bibnamefont
  {Loss}}\ and\ \bibinfo {author} {\bibfnamefont {D.~P.}\ \bibnamefont
  {DiVincenzo}},\ }\href@noop {} {\bibfield  {journal} {\bibinfo  {journal}
  {Physical Review A}\ }\textbf {\bibinfo {volume} {57}},\ \bibinfo {pages}
  {120} (\bibinfo {year} {1998})}\BibitemShut {NoStop}%
\end{thebibliography}%
	
\end{document}